\documentclass[12pt,a4paper,english]{article}
\usepackage[T1]{fontenc}
\usepackage[latin2]{inputenc}
\usepackage[english]{babel}
\usepackage{amsmath,amssymb,pstricks}
\usepackage{amsfonts}
\usepackage{indentfirst}
\usepackage[dvips]{graphicx}

\begin{document}

\sloppy


\begin{center}
\Large{ \bf How knots influence properties of proteins}
\end{center}

\begin{center}

\bigskip

Joanna I. Su{\l}kowska$^{1,2}$, Piotr Su{\l}kowski$^{\,3,4}$, P. Szymczak$^5$ and Marek 
Cieplak$^1$

\bigskip

\bigskip

\emph{$^1$Institute of Physics, Polish Academy of Sciences, \\ Al. Lotnik\'ow 32/46, 02-668 
Warsaw, Poland}

\medskip

\emph{$^2$CTBP, University of California San Diego, Gilman Drive 9500, La Jolla 92037}

\medskip

\emph{$^3$Physikalisches Institut der Universit{\"a}t Bonn and Bethe Center for Theoretical 
Physics, 
\\ Nussallee 12, 53115 Bonn, Germany}

\medskip

\emph{$^4$So{\l}tan Institute for Nuclear Studies, Ho\.za 69, 00-681 Warsaw, Poland}

\medskip

\emph{$^5$Institute of Theoretical Physics, University of Warsaw, Ho\.za 69, 00-681 Warsaw, 
Poland}

\bigskip

\bigskip

\smallskip
 \vskip .6in \centerline{\bf Abstract}
\smallskip

\end{center}

Molecular dynamics studies within a coarse-grained structure based model
were used on two similar proteins belonging to the transcarbamylase family 
to probe the effects in the native structure of a knot.
The first protein, N-acetylornithine transcarbamylase, contains no knot whereas
human ormithine transcarbamylase contains a trefoil knot located deep within the sequence.
In addition, we  also analyzed a modified transferase with the knot removed
by the appropriate change of a knot-making crossing of the protein chain.
The studies of thermally- and mechanically-induced unfolding  processes
suggest a larger intrinsic stability of the protein with the knot.

\bigskip

\textsf{knots | proteins | force-induced stretching  | molecular dynamics | AFM}

\bigskip

\newpage

\section{Introduction}

After the first discovery of knotted proteins \cite{Taylor_2000}, considerable 
attention has been devoted to the identification of the types of knots that are present in the 
protein structure base 
\cite{Virnau,Taylor_2007}. One interesting subclass identified contains more subtle  
topological configurations called $slipknotted$ proteins \cite{Yeates_2007}.
While structure-based analysis are become increasingly available, there are few studies 
describing 
the dynamical properties of knotted proteins. 
Simulations of the folding of the small knotted protein 1j85, combined with  
experimental results \cite{mallam,mallam0} led Wallin et al. \cite{shak} to propose that 
non-native contact interactions  
are necessary to fold a protein into a topologically non-trivial conformation. 
Interestingly, in studies of the tightening of knots  
under stretching at constant velocity the knots were found to jump 
between a set of characteristic sites, typically endowed with a large 
curvature, before arriving at the final fully tightened conformation \cite{Sulkowska_2008}. 
These results are in direct contrast to the well studied case of knots in 
homopolymers which tend to diffuse smoothly along the chain and then 
eventually slide off \cite{Metzler_2006}. 

It remains unclear whether knots are responsible for any biological functions 
or just occur accidentally. One noteworthy suggestion posed is that  
they provide additional stability necessary for maintaining the global fold 
and function under harsh conditions\cite{Taylor_2007}. 
Indeed, RNA methylotransferase derived from thermophilic bacteria appears 
to require knots for optimal function \cite{Nureki}. 
Consistent with the functional hypothesis, knots are usually found within catalytic domains 
of enzymes \cite{Taylor_2007}. Sometimes they encompass active sites \cite{Taylor_2007} 
where additional stability or rigidity could enhance catalysis  
when substrates are bound \cite{Virnau,Taylor_2007}. 

Thus it is important to understand how the presence of a knot 
may influence the properties and behavior of proteins in solution. 
In this paper, we consider three proteins within the same superfamily which are almost identical 
and differ by  the presence or absence of a topological knot. 
Two of the proteins are N-acetylornithine transcarbamylase AOTcase (the PDB code 1yh1) 
and ormithine transcarbamylase OTCase (1c9y) 
where the former has a knot \cite{Virnau} and the latter does not contain this topological 
feature.   
The third structure is a synthetic construct made from 1yh1 by redirecting the 
backbone so that the knot is removed. This system will be referred to as 1yh1$^*$. 
We focus on thermal and mechanical unfolding processes in these systems and compare  
the properties of these proteins {\it in silico} within  
a structure-based coarse-grained model as implemented in  
\cite{Cieplak_2004,Cieplak_2004a,Sulkowska_2007b}.  
In particular, we consider  AFM-imposed stretching at constant  
velocity and at constant force and determine the characteristic times for the thermal unfolding 
and the folding temperature. 
In all cases, the  knotted protein is more stable to unfolding.  
We compare these results with those observed for the sidechain disulphide bridged knots.

\section{The proteins studied}

The proteins 1yh1 (discussed in \cite{Shi_1yh1}) and 1c9y 
(discussed in \cite{Shi_1c9y}) belong to the transcarbamylase superfamily which is essential 
for arginine biosyntesis \cite{Morizono}.  The structures are nearly identical except that 
1yh1 contains a knot in its native structure whereas 1c9y does not \cite{Virnau}. 
The presence or absence of the knot seems to be responsible for the observed differences in 
enzymatic 
properties of the two proteins.

\begin{figure}
\begin{center}
\includegraphics[width=0.41\textwidth]{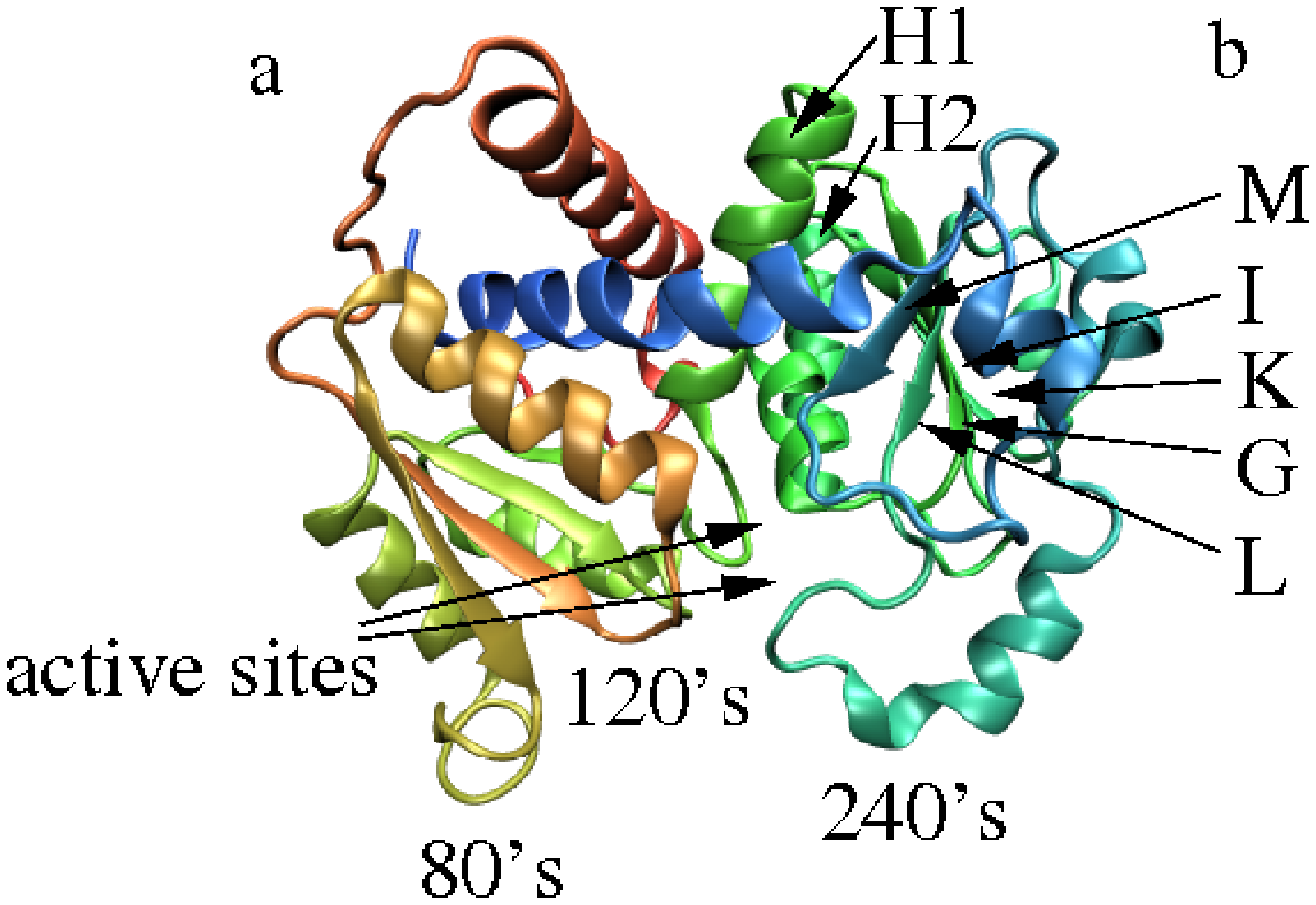} $\qquad$
\includegraphics[width=0.22\textwidth]{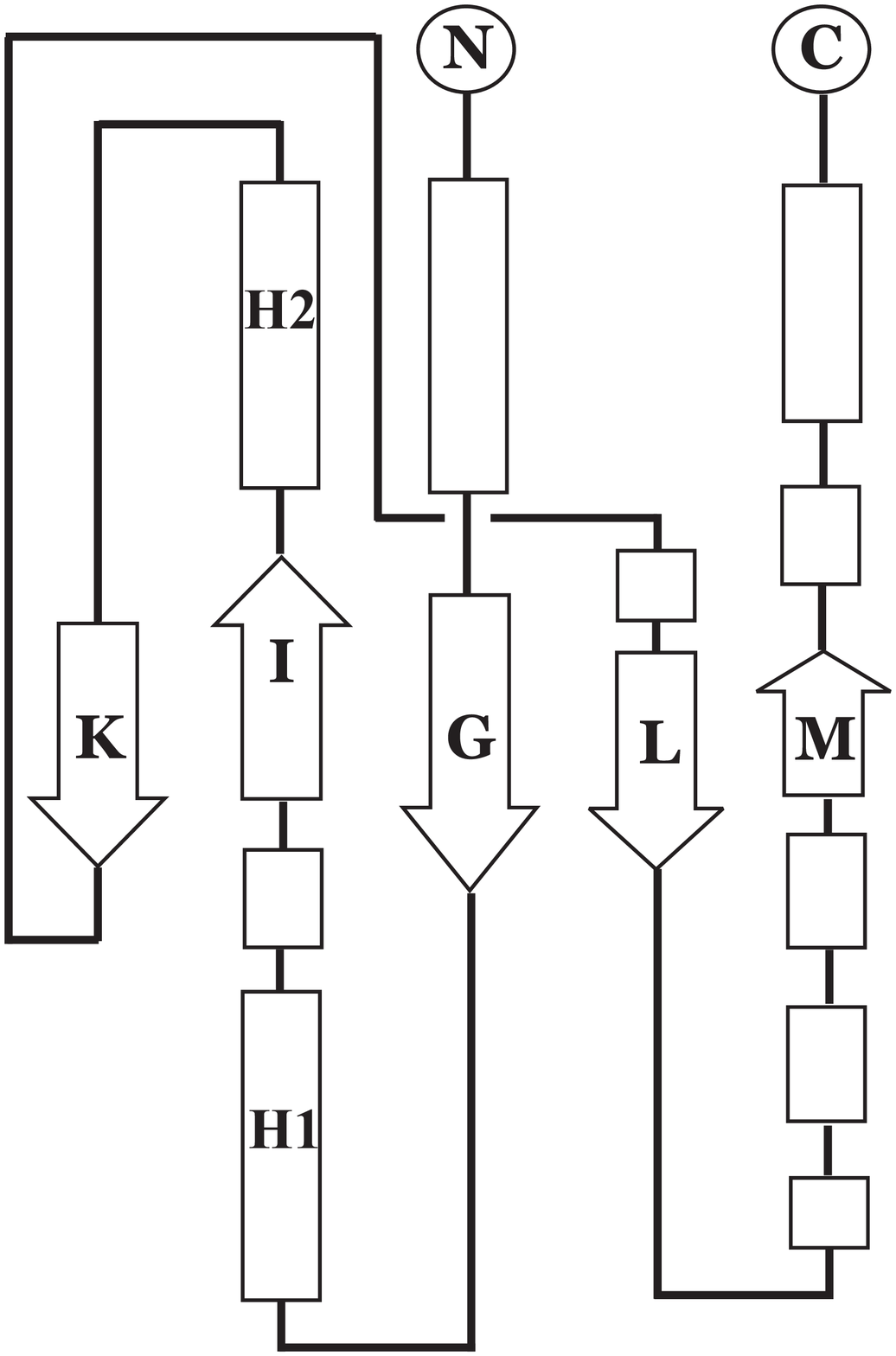} \\
\vspace{0.9cm}
\includegraphics[width=0.38\textwidth]{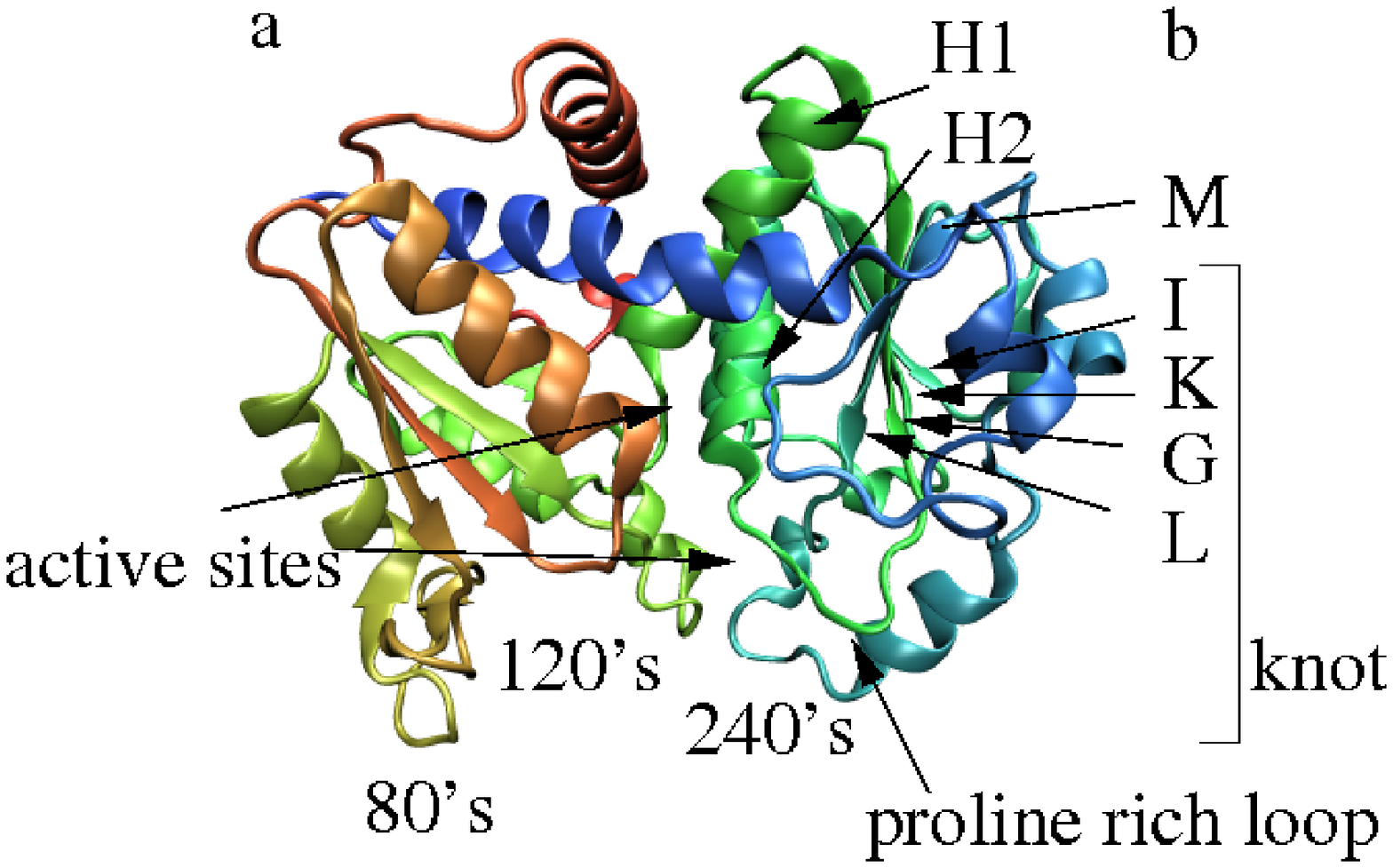} $\qquad$
\includegraphics[width=0.22\textwidth]{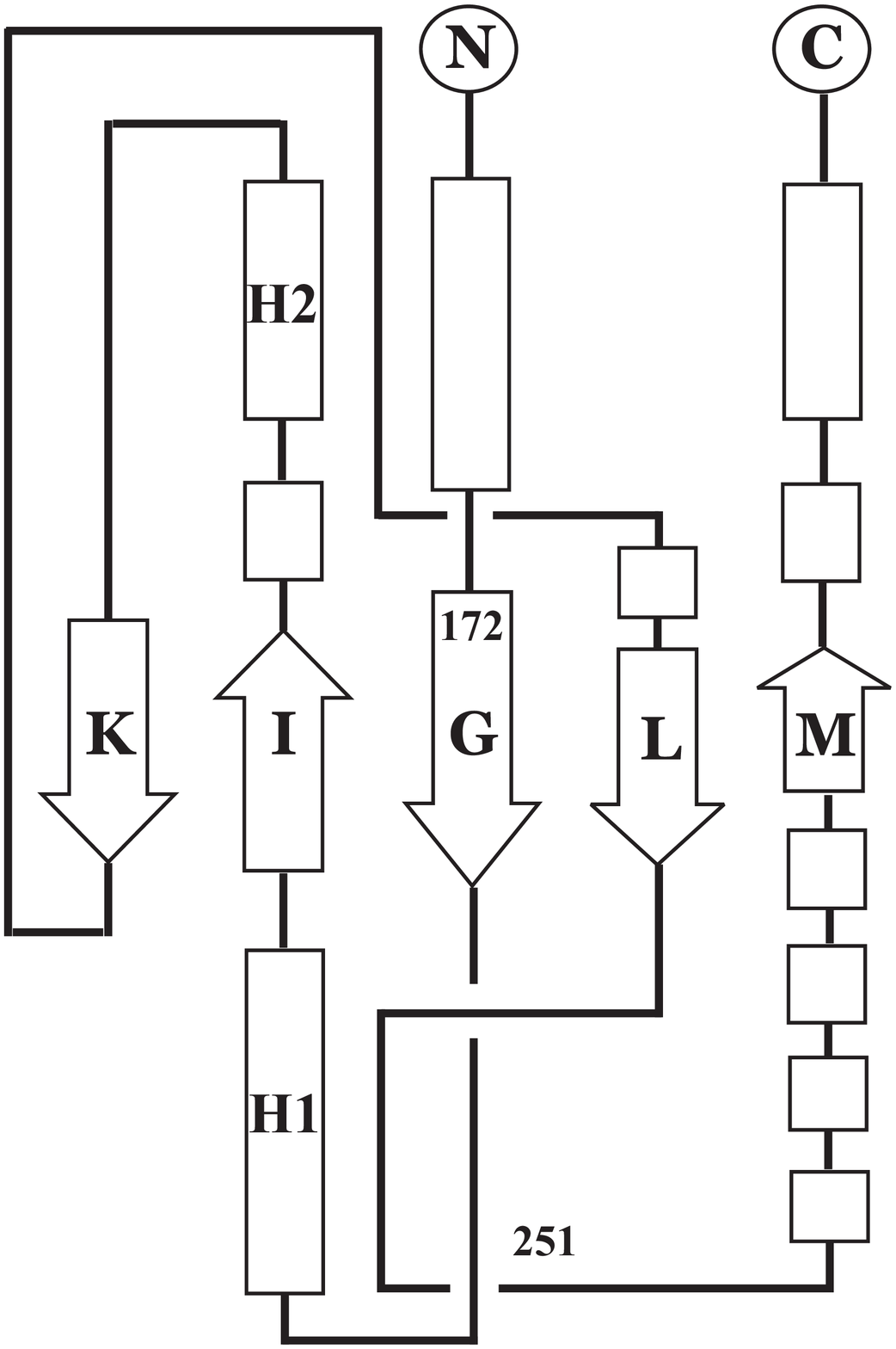}
\vspace{0.5cm}
\caption{Left: the cartoon representation of the unknotted 1c9y (top)  and knotted
1yh1 (bottom) proteins. Both consists of two $\beta$-domains, denoted as $a$ and $b$.
Right: domain $b$ is topologically trivial in 1c9y (top), while knotted in 1yh1 (bottom).
The arrows indicating the active sites are arranged in such a way that the upper (lower)
arrow corresponds to the first (second) active site. The knot in the native state in 1yh1
extends between amino acids 172 and 251 (whose locations are denoted
in a schematic figure on the right).
} \label{figbialka}
\end{center}
\end{figure}

Both proteins 1yh1 and 1c9y comprise two main $\beta$ domains 
denoted as $a$ and $b$, linked by the two interdomain helices 
(Figure \ref{figbialka}). 
The "weaving pattern" in domain $b$ 
is the structural feature that distinguishes the two proteins topologically. 
The $a$ domain in 1yh1 incorporates $\beta$ strands A(40-45), B(66-70),  
C(79-80), D(93-94), and E(108-112)  
whereas the $b$ domain -- strands G(172-177), I(202-206), K(232-236), L(248-252), 
and M(290-292) which create two main $\beta$ sheets. 
Both $\beta$ sheets are surrounded by many $\alpha$ helices. 
Strands C and D are quite short, but they create an extended loop 
around site 80, denoted the 80's loop, 
which is shorter in 1c9y  where strands C and D are missing altogether.

The sequential positions at which the knot begins and terminates are denoted by $n_1$ and $n_2$. 
These positions 
are determined by the KMT algorithm (see Materials and Methods).  
We use this algorithm at every step of our simulations, 
thereby obtaining the trajectories of knot's ends in the sequential space, 
such as those shown in the bottom panels of Figure \ref{fig-F-d}.  
The trefoil knot structure present in 1yh1 extends between amino 
acids $n_1$=172 and $n_2$=251 making it a relatively rare example  
of a "deep" knot since it is positioned relatively far from the termini of the protein.  
The knot encompasses almost the entire domain $b$, i.e. four $\beta$ strands G, K, L, I 
and two nearby $\alpha$ helices which we denote by H1 and H2 (also present in 1c9y). 
An important structural difference between 1yh1 and 1c9y is the presence of the  
proline-rich loop (181-183) in the former, a main building block for the knot-making crossing 
of the protein chain.

The two enzymes, OTCase and AOTCase participate, in the arginine biosynthetic pathway, 
however, the presence of the knot in AOTcase makes the corresponding pathway 
distinct \cite{Shi_2006}. 
Both proteins contain two active sites -- the first binds carbonyl phosphatase CP 
whereas the second site (which is modified by the knot structure) 
binds either N-acetylornithine  or L-ornithine, 
in the case of 1yh1 and 1c9y, respectively. The second site 
facilitates the chemical reaction with carbamyl phosphate to form 
acetylcitrulline or citrulline, correspondingly \cite{Shi_1yh1,Shi_1c9y}. 
We use the notation for the active sites introduced in \cite{Shi_1yh1,Shi_1997}, 
as shown in Figure \ref{figbialka}. 
The first active site, located between the two domains, 
is the same in the two proteins \cite{Shi_1yh1}. 
However, in 1yh1 the second active site is formed by Glu144 (within the extended 80's loop), 
Lys252 (from 240's loop), and the proline rich loop (which creates the knot). 
On the other hand,  in 1c9y the second active site  
is localized near the 240's loop \cite{Shi_1yh1,Shi_1c9y}. 
Thus the proline rich loop in 1yh1 does not allow the formation of contacts between a 
ligand and the 240's loop (which is possible in 1c9y) and leads to a different functional and 
topological 
motif.

The OTCase pathway shows ordered two substrate binding with large domain movements, 
whereas in the AOTCase pathway the  
two substrates are bound independently with small reordering of the 80's loop, 
small domain closure around the active site, and a small translocation  
of the 240's loop \cite{Shi_2006}.  
Thus it seems that the knot plays two roles here: it changes the environment 
for the second substrate N-acetylcitrulline binding, and -- as shown in this paper -- 
makes the structure more stable. As a result, the 
functional and thermodynamic properties of the fold are affected by the presence of the knot.

The proteins 1yh1 and 1c9y have similar numbers of native contacts 
(as determined based on the van der Waals radii of heavy atoms \cite{Tsai}),  
943 and 919 respectively, so any differences in properties must arise primarily from
rearrangements in connectivities in the contact map  .

The folding, thermal and mechanical properties of these two proteins 
have not been compared up to now, mostly because the structure of AOTCase 
has not been known until recently and because the presence of the knot makes 
experimental data harder to interpret. However, some experimental work has been performed 
on them as detailed in \emph{Appendix}.

We have also analyzed a modified 1yh1, in which the knot was removed by reversing the 
crossing created by the parts of the backbone contained between amino acids 
175-185 and 250-260. The cutting and pasting of these two parts of structure was done 
using all-atom techniques described in \cite{Gront_2007,Gront_2008}. 
The resulting structure 1yh1$^*$ has the same unknotted topology as 1c9y while 
it has 14 fewer contacts than the original 1yh1. This procedure affect the contact 
in the  vicinity of the original knot-making crossings while leave the global contact map 
intact.
The idea of rebuilding proteins 
to test their properties is a familiar one -- another interesting example of such 
protein engineering was discussed recently in \cite{Gosavi_2008}.

\section{Resistance to mechanical stretching}

One way to probe the stability of a biomolecule is to perform mechanical 
manipulations on it, such as stretching. The corresponding experimental 
data on the two proteins are not yet available, thus we have resorted to computer modeling. 
We consider the case in which the termini are connected to elastic springs. 
The N-terminal spring is anchored to a substrate and the 
C-terminal spring is pulled either at a constant velocity, $v_p$, or at 
constant force.

\subsection{Stretching at constant velocity}

\begin{figure*}
\begin{center}
\includegraphics[width=0.91\textwidth]{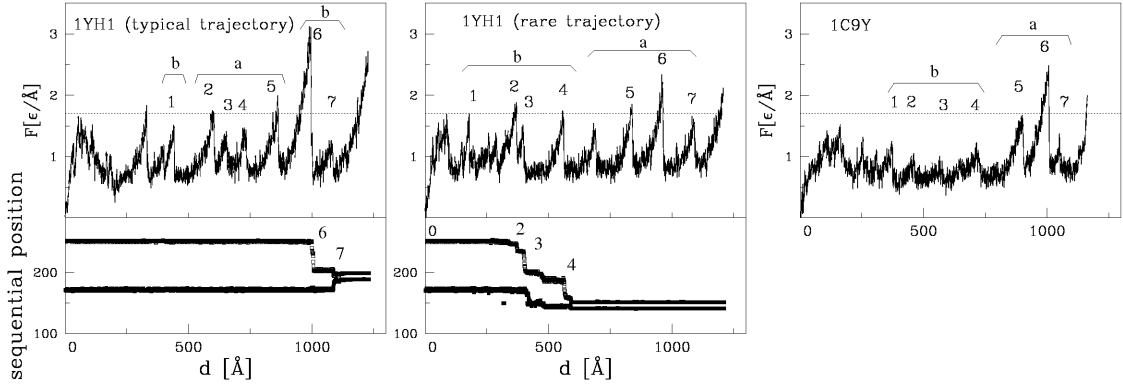}
\caption{Top: Unfolding curves of force versus protein length $F(d)$ stretched at
constant velocity $v=0.005$ {\AA}$\tau$ as explained in the text.
The horizontal dotted line indicates a
reference of $F$=1.7 $\epsilon /${\AA} corresponding to the hight of many
of the force peaks. It is drawn to facilitate panel-to-panel comparisons.
The initial force peaks do not relate to the $a$ and $b$ domains.
The remaining force peaks are labeled 1 through 7 except that in the middle
panel there is an extra peak between 4 and 5 corresponding to shearing of
helices that are coupled to the $a$ domain
In each case, the force peak labeled as 1 arises due
to shearing of the L strand against the M strand.
Table I lists which contacts break (i.e. $r_{ij} \; > \; 1.5 \sigma _{ij}$)
at the remaining the remaining peaks.
Bottom: Sequential movement of knot's ends during the knot tightening process
corresponding to the trajectory shown above.
 } \label{fig-F-d}
\end{center}
\end{figure*}

In this mode of manipulation, one monitors the force of resistance to 
pulling, $F$, as a function of the pulling spring displacement, $d$. 
We usually take $v_p$=0.005 {\AA} /$\tau$ which is about 100 times faster than 
typical experimental speeds. Results obtained for $v_p$=0.001 {\AA} /$\tau$ are 
found to be similar. In the absence of thermal fluctuations a single unfolding trajectory 
is followed. At finite temperatures, however, differences between various 
trajectories arise. Usually, these differences are small. Such is the case for the 
unknotted 1c9y for which a typical $F(d)$ trajectory is shown in the rightmost 
panel of Figure \ref{fig-F-d}. However, for 1yh1 we identify two  
distinct pathways. The major pathway is shown in the middle panel of 
Figure \ref{fig-F-d} and the alternative pathway in the leftmost panel. 
In fact, that pathway is quite rare: it has been found just once in 
fifty trajectories. The locations of the knot ends 
during stretching are displayed in the lower regions of the two panels. 
The immediate conclusion is that the knotted protein 1yh1 is typically more 
resistant to stretching than 1c9y since the maximum force peak, $F_{max}$, is about 3.3 
compared to 2.6 $\epsilon/${\AA} (2.9 
and 1.7 $\epsilon /${\AA} for $v_p=0.001$ {\AA}/$\tau$), with the energy scale $\epsilon$ 
as defined in the Materials and Methods. It is only for the rare trajectory that the 
values of $F_{max}$ for the two proteins are nearly the same,
but even then the unfolding pathways are distinct as 
evidenced in Table I.  Based on the data presented in ref. 
\cite{Sulkowska_2008d,Sulkowska_2008a}, the unit of 
force, $\epsilon /${\AA}, used here should be of order of 70 pN. 
There are uncertainties in this estimate (of order 30pN), but the important observation is 
that we compare similar proteins with a similar effective value of the $\epsilon$.

Table I shows that the unravelling of both proteins proceeds along different pathways. 
Unfolding of the unknotted 1c9y starts from domain $b$ (which is stabilized by 
the knot in 1yh1) and once this domain is fully unravelled the unwinding 
of domain $a$ follows. In the knotted 1yh1 also the domain $b$ begins to unfold first. 
However, in the typical pathway, its unfolding stops relatively soon, just after the  
strands L and M are pulled apart since the next step would disarrange the knot. Instead, 
the domain $a$ is unfolded first, and only then the process of knot tightening 
begins.

We note that the first broad peak for each trajectory from Figure 
\ref{fig-F-d} corresponds to the shearing motion between two domains, which are connected 
by two alpha helices. 
It has been established experimentally \cite{Shi_2006} that the interdomain 
interactions in 1yh1 are slightly stronger than in 1c9y and are mainly hydrophobic, which 
is consistent with our observation that the first peak in 1yh1 is higher than in 1c9y. 
Also the origin of the main force peak is different in the two proteins: in $1yh1$ (typical 
pathway) it coincides with knot tightening within domain $b$, 
which is accompanied by shearing of the $\beta$ strands G+I, G+L, I+K. 
In contrast, in 1c9y the main peak is associated with 
shearing the $\beta$ strands A+B, A+E within domain $a$. On the other hand, 
the rare unfolding pathway of 1yh1 shares many features with 
that of 1c9y. Nonetheless, due to the presence of the 
knot, pulling the strands in domain $b$ apart involves a higher force 
than  in 1c9y (where the $b$-domain related peaks appear at distances 400-700 \AA). 

We now consider constant speed stretching of the synthetic protein 1yh1$^*$. 
Two alternative stretching pathways are also observed in this case, 
as shown in Figure \ref{fig-F-d2}. 
The typical pathway (8 out of 10 trajectories) yields 
$F_{max}$ of just below 2.5 $\epsilon$/{\AA} which is smaller than 
$F_{max}$ for the typical pathway in 1yh1 by $\sim 0.5$ $\epsilon$/{\AA}. 
The minor pathway yields 
$F_{max}$ which is smaller by $\sim 0.2$ $\epsilon$/{\AA} than the corresponding value in 
1yh1. This lowering in the value of $F_{max}$ clearly points to the dynamical 
significance of the knot. 
In the typical case, the unfolding process is found to proceed in the same way as in 
the unknotted 1c9y: domain $b$ unfolds first, followed by $a$.  
On the other hand, in the alternative trajectory, 
domain $b$ first unfolds partially, then complete unfolding of $a$ follows, and  
only then unravelling of $b$ is completed. This pathway is  
analogous to the typical unfolding of the original knotted 1yh1. However, 
it is the unfolding of domain $a$ (and not $b$) which is  
responsible for the main force peak in  1yh1*. The corresponding  value of 
$F_{max}\simeq 2.4\epsilon$/\AA $\ $  
is close to the $F_{max}$ observed for the unknotted 1c9y 
(where it also arises from unfolding of domain $a$). All of these observations 
indicate that the dynamical  
differences between 1yh1 and 1c9y can indeed be attributed to the presence 
or absence of a knot in the former. 

We now discuss the process of knot tightening and focus on the knotted 1yh1. 
Similar to what has been found in other proteins with knots 
\cite{Sulkowska_2008}, the knot ends in 1yh1 make sudden 
jumps to selected metastable positions. 
Figure \ref{fig-F-d} shows that those jumps are correlated with 
the force peaks corresponding to unfolding events in domain $b$. 
In the typical case (the left panel), 
the knot moves to one of the metastable places at 
$d\sim$ 1000 \AA $\ $ (where $F$ becomes $F_{max}$), which is followed by 
tightening of the knot, usually in two additional steps. 
As shown in \cite{Sulkowska_2008} the set of possible sites at which an end may land 
corresponds to the sharp turns in the backbone (usually with proline or glycine).
In our case, 
the sites  Gly-200, Pro-210 and Gly-230 are found to be the most likely choices.
It is interesting to note that for the rare pathway in the set of possible sites at
which an end
1yh1 (middle bottom panel in Figure \ref{fig-F-d}), 
the knot first moves from the native position (172,251) 
to (Val-140, Gln-151). The new knot end positions are close to Pro-139 and 
Pro-149 which makes this location stable.
In proteins comprising less than 151 amino acids, $F_{max}$  tends to 
arise at the beginning of the stretching process \cite{Sulkowska_2007b}. 
Here, however, the proteins are large and 
adjust to pulling by first rotating to facilitate unfolding of other parts in their 
structure, and only then by unraveling the harder knotted part.

We also analyzed stretching of tandem linkages of the proteins. 
Two proteins 1c9y linked together are found to unravel in a serial fashion. 
This is not the case, however, for two domains of 1yh1.  
When the unfolding process in one domain reaches the knot region, the other domain starts to 
unfold. In the final stages both knots tighten simultaneously.

\subsection{Comparison between the effects of knots and of disulphide bridges}

In the current study we demonstrate that knots provide extra mechanical stability to proteins. 
Thus, one may think of knots as acting analogously to disulphide bridges 
between cysteins. Like knots  (with the exception of a situation 
in which pulling unmakes the knot), 
the disulphide bridges cannot be removed from proteins by stretching.  
However, unlike knots, they cannot slide along 
the sequence. Furthermore, the bridges can be weakened through application 
of the reducing agent DTT as in refs. \cite{Discher_2001,Discher_2004}.  
As a theoretical analogue of the cysteine knot-containing hormons  
studied by Vitt et al. \cite{Vitt}, we consider a hypothetical mutated  
version 1c9y in which amino acids at sites 195 and 265 (one could also 
consider 194 and 262) are replaced by cysteins. The resulting disulphide  
bridge linking the two sites would close a knot-like loop. 
The presence of a disulphide bridge can be imitated by strengthening 
the amplitude of the Lennard-Jones contact potential to 
$\epsilon _{ss}=\zeta \epsilon$. We consider  $\zeta =20$ 
which makes the bridge essentially indestructible.

\begin{figure}
\begin{center}
\hspace{-1cm}
\includegraphics[width=0.91\textwidth]{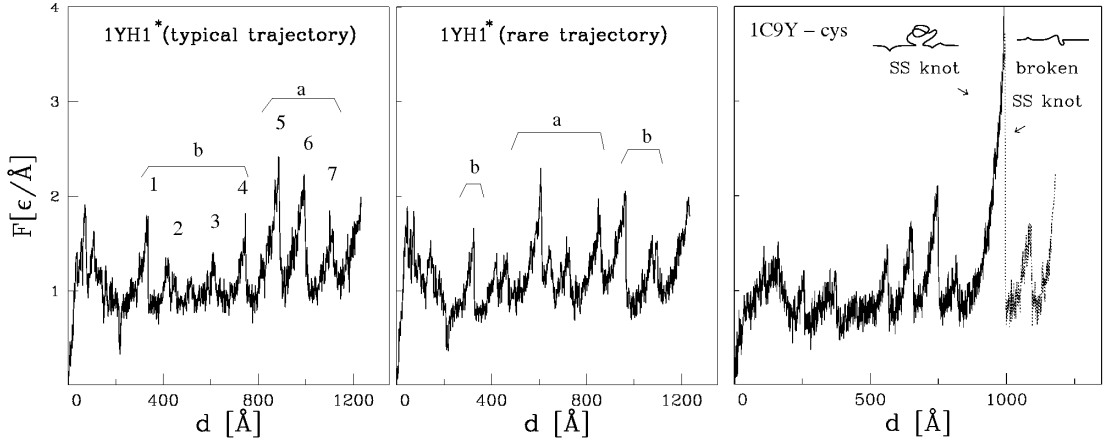}
\caption{$F(d)$ curves for the synthetic 1yh1* without the knot (left and center panel).
and for the mutated 1c9y with
the disulphide bridge (right panel). In the latter, the solid line corresponds to $\zeta$=
20
and the dotted line to $\zeta$=10.
} \label{fig-F-d2}
\end{center}
\end{figure}

The rightmost panel of Figure \ref{fig-F-d2} shows that the resulting $F(d)$ pattern is 
quite similar to the typical trajectory for 1yh1 shown in the left panel 
except for a diverging force peak towards the end of the process.  
One can endow the disulphide bond with more pliancy by reducing $\zeta$  
to the value of 10 and thus allowing for the continuation of the stretching process 
(the dotted line in Figure \ref{fig-F-d2}). The corresponding sequence of the rupture  
events (L+M, followed by A+B, then A+E, then E+F, then G+L, G+I, H1, H2, and finally I+K) 
is different than any of 1yh1 unfolding pathways ( Table I). 
However, the order of events seems closest to the typical  
trajectory found for 1yh1: partial unwinding of domain $b$, followed by 
unwinding of $a$ and then returning to unravel $b$. 
We conclude that even though the disulphide bridges act dynamically 
similar to the knots there are also differences in the details.

\subsection{Stretching at constant force}

\begin{figure*}
\begin{center}
\includegraphics[angle=-90,width=0.91\textwidth]{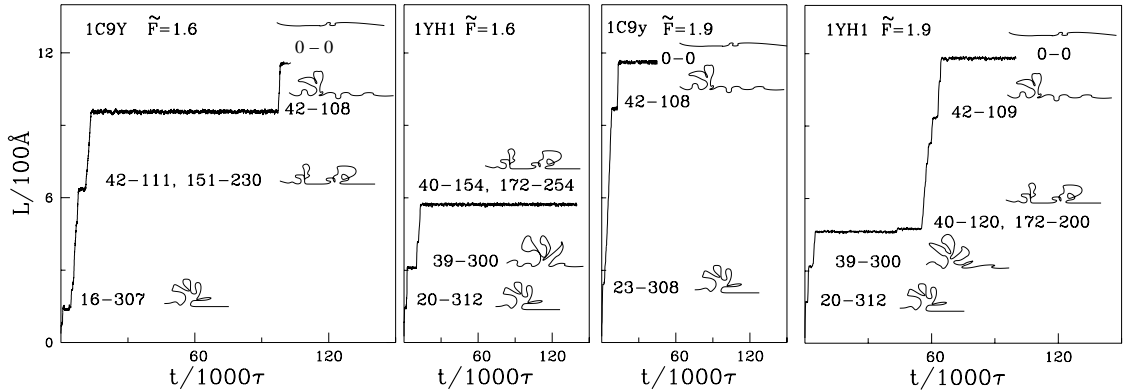}
\caption{The time dependence of the end-to-end distance when stretching by
constant force for the indicated values of the force. The left panels refer
refer to the unknotted protein and the right panels to the knotted one.
Schematic pictures of the conformations corresponding to the metastable state
are displayed on the right hand side of each panel where the $a$ and $b$
domains are depicted as blobs.
} \label{cforce}
\end{center}
\end{figure*}

\begin{figure}
\begin{center}
\includegraphics[width=0.5\textwidth]{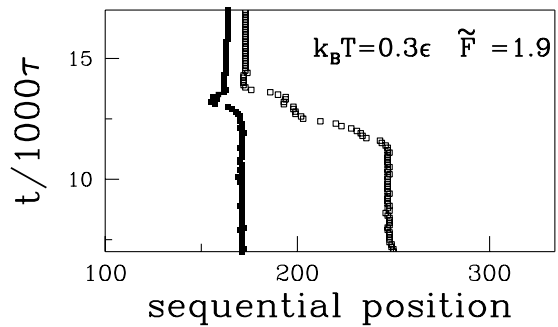}
\caption{A typical trajectory of knot's end locations for stretching at constant force. }
\label{fig-cforce-knot}
\end{center}
\end{figure}

The dynamical differences between the knotted and unknotted proteins 
should  also be visible when performing stretching at a constant force, F. 
In this mode of manipulation, one monitors the end-to-end distance, $L$, 
as a function of time as illustrated in Figure \ref{cforce} for selected 
trajectories. In each trajectory, $L$ varies in steps indicating transitions 
between a set of metastable states that depend on the applied force. 
For $\tilde{F} < 1.7$  (where $\tilde{F}$ denotes $F$ in units of $\epsilon$/{\AA}),
domain $b$ in 1c9y gets unraveled first while domain $a$ remains intact.
Once the system reaches $L$ which is just above 900 {\AA}, it stays 
at this extension indefinitely. For larger forces, the $b$ domain also unravels and the ultimate 
value of $L$ reached is $\sim$1200 {\AA}.
The pathways observed for the  knotted protein 1yh1 the are rather different. 
For $\tilde{F} < 1.7 $ neither domain $a$ nor $b$ unfolds 
indicating again the stabilizing role of the knot. It is only the 
remaining parts of the structure that unravel leading to the largest $L$ of 600 {\AA}.
For $\tilde{F}$ between 1.7 and 1.9, two pathways are possible. 
In the first one, domain $a$ remains nearly intact while domain $b$ gets unfolded, 
leading to tightening of the knot and to a maximum value of $L$ of 950 {\AA}. 
This situation is analogous to the one found for 1c9y.  
In another pathway, the $a$ domain unfolds first, but again full 
extension of the chain is not achieved. 
For $\tilde{F} > 1.9 $ the $b$ domain always is always the first to unfold. 
The related movement of knot's ends 
are shown in Figure \ref{fig-cforce-knot}. The knot tightening process looks similar
to the one observed in the \emph{rare} trajectory for the constant velocity 
stretching (Figure \ref{fig-F-d}, middle panel). 
In this case, domain $a$ eventually unfolds, leading to full extension of the chain.
For $\tilde{F} > 1.9$, the scenarios of unfolding for 1yh1 and 1c9y 
are almost identical (except for the breakage of C+D bonds, which are absent in 1c9y) 
and are summarized in Table I. 
However the time intervals between consecutive steps are typically longer for 1yh1,  
indicating a slower unfolding process.
An analysis of the results of stretching with constant velocity lead us to expect
an interesting behavior for the results for $F \approx F_c$=1.7 $\epsilon/${\AA}, as
the values of  $F_{max}$ (corresponding to domain $b$) for 1c9y  seen in Figure
\ref{fig-F-d} are much lower than $F_c$, while for 1yh1 some 
of them are above $F_c$ (both in the typical and rare trajectories). 
The characteristic value $F_c$ is indicated in Figure \ref{fig-F-d} by the horizontal dotted
line. Indeed, for stretching with a force $\ge F_c$, we do not observe
any steps in the curves $L(t)$ that are related to peaks 1-4 for 1c9y 
(Figure \ref{fig-F-d}, the right panel). 
On the other hand, we still observe such structures (corresponding to the highest among peaks
1-5 in Figure \ref{fig-F-d}, the left and middle panels) during 
stretching of 1yh1 with $F=F_c$ (and slightly higher). 
Such a behavior is seen in Figure \ref{cforce} for $F=$1.9 $\epsilon$/{\AA}.

We also analyzed in detail an example of a constant force 
pathway for 1yh1 for $F > 1.9$ $ \epsilon/$\AA $\ $ 
(see Figure 5 and Suppl. Mat.) The scenarios of events reported here are consistent 
with the prominent role of the sharp structural turns in the dynamics of 
knot's ends \cite{Sulkowska_2008}. In particular, when 
the knot is tightened, its right end moves across several pinning centers 
comprising the turn between $\beta$-strand K and the small helix H2, and 
the sharp turn at Pro-210, slowing down at successive pinning centers.  
Each slowing down manifests itself as accumulation of points along the tilted 
interval in Figure 5. 
 


\begin{figure}
\begin{center}
\includegraphics[width=0.6\textwidth]{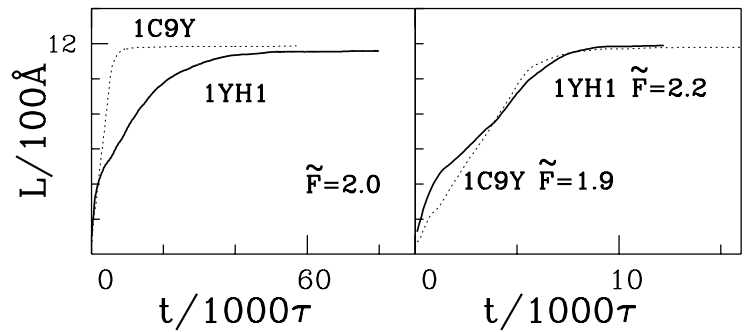}
\caption{The average end-to-end distance as a function of time for the
forces indicated.}
\label{aver_force}
\end{center}
\end{figure}

So far we have discussed the differences between 1c9y and 1yh1 
as seen at the level of single stretching trajectories. These differences 
are also visible after averaging over many trajectories, 
as demonstrated in Figure \ref{aver_force}. 
In particular, we find that for  
$\tilde{F}= 2.0$ (shown in the bottom panel), 
which allows for the full extension in both proteins, 1yh1 takes longer 
to unfold than 1c9y. 
However, for forces higher than $2.3 \; \epsilon$/{\AA} the differences in the
averaged trajectories are minor. 
The top panel shows that in order to match the time scale of unfolding 
1c9y at $\tilde{F}=1.9 $ in 1yh1 one has to enhance the value of 
$\tilde{F}$ to 2.2.

\begin{figure}
\begin{center}
\includegraphics[width=0.55\textwidth]{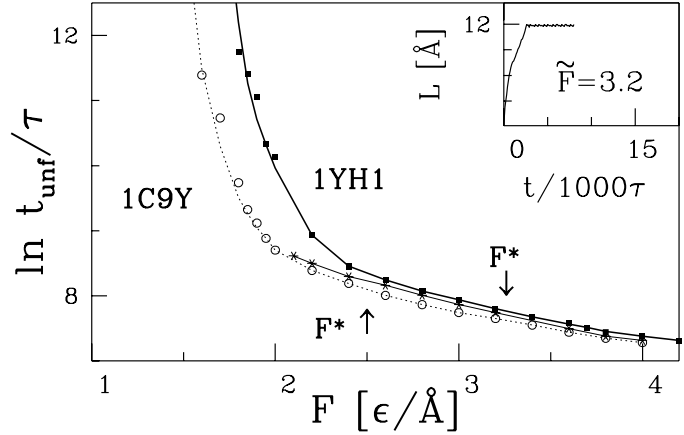}
\caption{The unfolding times $t_{unf}$ as a function of the force applied.
The solid fat line (with squares) and solid fine line (with asterisks)
are respectively for 1yh1 and 1yh1*, the dotted line (with circles) for 1c9y.
Inset: for $\tilde{F}$=3.2 the protein is stretched instantaneously, without
formation of any metastable states, and with small trajectory-to-trajectory variations.}
\label{fig-tunf-F}
\end{center}
\end{figure}

One can quantify the time scales of the force induced unfolding by determining 
the mean time, $t_{unf}$, needed to break all contacts with a sequential distance 
$|j-i|$ bigger than a threshold value $l_c$  (a somewhat different  
criterion has been used in ref. \cite{Szymczak_2006}), 
see also a related study by Socci et al. \cite{Socci}.  
The smaller the $l_c$, the longer the corresponding $t_{unf}$. In practice, 
we have found it feasible to take $l_c=8$. As shown in Fig.~\ref{fig-tunf-F} the resulting  
unfolding times, $t_{unf}(F)$, are longer for 1yh1 than for 1c9y,  
which is another manifestation of the higher stability  of 
the knotted protein. The stability of 1yh1 is significantly reduced upon replacing 1yh1 by 
its synthetic variant 1yh1$^*$. Figure 7 also indicates the values of $F^*$ -- a force  
above which the unfolding commences instantaneously. Again, $F^*$ for 1yh1 is 
substantially higher than for 1c9y and 1yh1$^*$.

\section{Thermal stability}

\begin{figure}
\begin{center}
\includegraphics[width=0.55\textwidth]{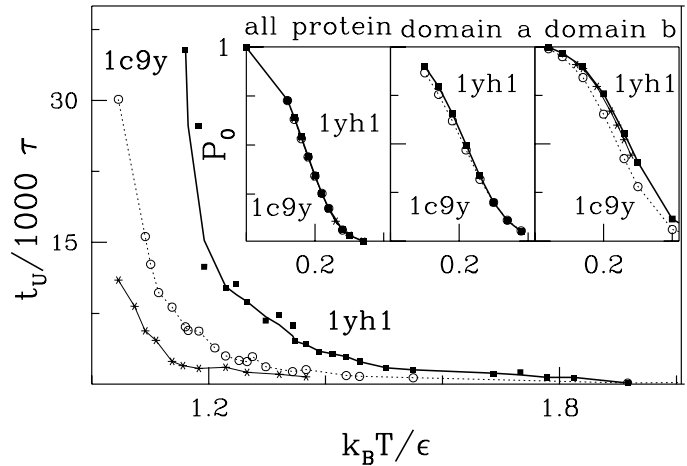}
\caption{The dependence of the median unfolding time on temperature.
The solid fat line (with squares) and solid fine line (with asterisks)
are respectively for 1yh1 and 1yh1*, the dotted line (with circles) for 1c9y.
Inset: The temperature dependence of the probability of preserving all the native
contacts in 1yh1 and 1c9y.}
\label{termo}
\end{center}
\end{figure}

We now consider unfolding via thermal fluctuations following the approach 
of Ref. \cite{Cieplak_2005b}. We define the unfolding time, $t_U$, as the median duration of a
trajectory that starts in the native state and stops when 
all contacts within $|j-i|\;> l_c$ get broken. For consistency with the 
mechanical studies, we choose $l_c=8$.  
The temperature dependence of $t_U$ for both proteins is shown in
Figure \ref{termo}. Clearly, for any given $T$, it takes substantially 
longer to unravel 1yh1 than either 1c9y or 1yh1$^*$. For instance, at $k_BT/\epsilon$=1.3 
the ratio of $t_u$ between 1yh1 and 1c9y is about 2.

It should be noted that the mere fact that the contacts with the sequential length larger 
than $l_c$ are broken does not necessarily mean that the knot itself  has loosened and 
become untied. In fact, according to our studies of thermal unfolding, the knotted 
proteins unfold in two steps: first the long-ranged contacts break and only then, at much 
longer time scales, the knot becomes undone. Thus the unfolding follows the 
$N \rightarrow UK \rightarrow U$ path, where 
$N$ stands for the native state, 
$UK$ for the unfolded knotted state and $U$ for the totally unfolded, unknotted state. 
Due to the topological constraints present in the $UK$ state, its entropy is considerably 
lower than that in $U$ state, thus the free energy difference between $UK$ and $N$ is much 
higher than that between $U$ and $N$, which leads to the increased stability of the native 
state.
Similar entropy-based strategies for increased stabilization are found in other 
topologically constrained proteins~\cite{Zhou}, e.g. in proteins with circular 
backbones, which  has been shown to be highly resistant to  
enzymatic, thermal and chemical degradation~\cite{Colgrave}. 
There is also another, energy-based reason for the increased stability of 1yh1 and, 
possibly, of other knotted proteins. Namely, non-trivial topology of a protein may lead to 
a more energetically favored conformational state. This is the case for the three proteins 
considered here: the knotted 1yh1 has the lowest native state energy. The native state 
energy of 1yh1$^*$, the unknotted counterpart of 1yh1, exceeds that of 1yh1 by 
$14 \epsilon$, whereas that of 1c9y is higher than 1yh1 by about $24 \epsilon$. Thus one of the
reasons why knots may be preferred in certain proteins is that they lead 
to deep native state minima. 

Apart from the higher stability of 1yh1, its longer unfolding times can
also be explained in terms of topological frustration 
\cite{Gosavi_2008,frust}. It arises when only a particular order of contact 
breaking allows the protein to unfold. When this order is incorrect, 
certain geometrical constraints arise which do not allow for 
unfolding, and some contacts are forced to form back again. Therefore 
a protein unfolds in a series of steps, also called a backtracking, 
which involve refolding and unfolding. The consequence of this geometric 
bias is an unusually long unfolding time. There are 
obvious geometrical constraints present in 1yh1 related to its 
knotted structure, so it is likely that its unfolding is dominated by 
topological frustration and takes more time than unfolding of 
unknotted 1c9y or 1yh1$^*$. A particular example of backtracking, which arises
in 1yh1 is presented in detail in \emph{Appendix}. 

To assess the magnitude of fluctuations around the native state we measured $P_0(T)$
defined as the fraction of time during which all native contacts are established for 
the trajectory starting in the native conformation. 
This quantity can be regarded as yet another measure of stability.
However, even though $P_0$ is calculated based on 
relatively long trajectories of $10^5 \tau$, 
this is still only a small fraction of the expected unfolding time in this range 
of temperatures. These trajectories are therefore not ergodic  
and probe vicinity of the native state basin. 
The results are shown in the inset of Figure \ref{termo}: the left panel shows the data for
entire length proteins whereas the middle and right panels are for the $a$ and $b$ domains, 
respectively.
In the right inset panel for domains $b$ (which contains the knot in 1yh1)  
the data points corresponding to 1c9y are shifted towards lower 
temperatures relative to 1yh1. A similar, however smaller shift towards lower temperatures 
is also observed for the synthetic 1yh1$^*$. On the other hand, data points for domain $a$ 
(the middle panel) and for the whole protein (left panel) are similar.  
Thus differences in $P_0$ are confined to domain $b$ 
and indicates a higher stability of domain $b$ in the knotted protein.

\subsection{Thermal untying of a knot in 1yh1 protein}

\begin{figure}
\begin{center}
\includegraphics[width=0.8\textwidth]{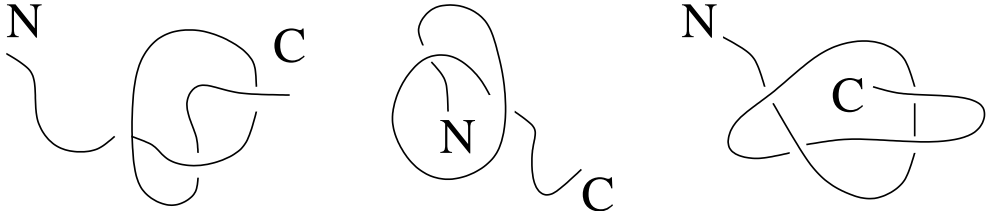}
\caption{Three possible ways of thermal untying of the knot. From the left to
the right: simple from the C terminus, simple from the N terminus
and through formation of a slipknot.
} \label{knot_rozwijanie}
\end{center}
\end{figure}

As mentioned above, untying of the knot involves much longer time scales than those
of long range contact breaking. However, the unknotting times
decrease with increasing temperature. 
Meaningful studies could be performed for $k_BT/\epsilon$=1.2 (and higher). 
We have found that the knot opens more readily on the side closer to the C terminus, 
while its N-terminus-side is more stable. This is in agreement 
with the results of \cite{Nterm} on the asymmetry of (slip)knots, and the fact that 
they arise much more often closer to the N-terminus. 
Examples of conformations 
corresponding to different ways of thermal untying of the knot are shown in Figure 
\ref{knot_rozwijanie}. For each terminus, there are two 
possibilities: either it is the last site to leave the knot or else
it is a leader that pulls the rest of the knotted loop behind it. 
The latter circumstance is known as a formation of a slipknot \cite{Yeates_2007}. 
It is interesting to note that application of a high temperature has been 
occasionally found to generate short lived additional (slip)knots, 
especially when the native knot has disappeared. 

As generally expected and demonstrated in ref. \cite{Cieplak_2005b} explicitly, 
the process of thermal unfolding is statistically reverse to folding. 
Thus the phenomena we observe for unfolding should also be observed  
in folding processes. This also suggests that the presence of the  
non-native attractive contacts is not necessary for formation of a knot.
Indeed, in a subsequent paper we show that proteins of nontrivial topology 
have the ability to fold to their native states without any non-native interactions 
involed. Such non-native contacts have been vital in folding simulations of 
Wallin et al. \cite{shak}. More details and particular examples concerning 
thermal untying and backtracking 
it may be accompanied by are presented in \emph{Appendix}.

\section{Discussion and conclusions}

We have considered three very similar proteins -- one with a knot and two 
without -- and determined their properties by using a coarse-grained  
native-geometry based model. Both mechanically and thermally, the protein with 
the knot has been found to be more robust and is characterized by  
longer unfolding times, which we attribute to topological and geometric
frustration. The larger robustness of 1yh1 relative to 1c9y relates to 
to the experimental results on OTCase and AOTCase pathways. 
The OTCase pathway shows the two-substrate binding involving large domain movements.
In this pathway, the order in which the substrates are bound is well defined. 
On the other hand in the AOTCase pathway, the two substrates are bound independently.
This process involves small reordering of the 80's loop, 
small domain closure around the active site, and a small translocation 
of the 240's loop \cite{Shi_2006}.

Other findings can be summarized as follows: 
The unknotted variant of 1yh1 has been found to behave like the unknotted 1c9y. 
Therefore we conclude that this is the nontrivial knot topology 
that is responsible for the peculiar properties of 1yh1. 
Disulphide bridges may imitate existence of knots to some degree. 
The kinetics of the knot untying and thus, by a reversal, the kinetics 
of formation of the knot may involve generation of other knots and slipknots.
According to \cite{Shi_1yh1}, the presence of the knot motif in  AOTCase 
affects the way the N-acetylcitrulline is bound to the second acive site and 
thus changes the arginine biosynthetic pathway. 
This observation can provide important information on potential targets for specific 
inhibition of bacterial pathogens. Such inhibitors would not affect the more common 
OTCase and thus provide a specific non-toxic method for controlling certain pathogens.

Taken together, these findings show that relatively small structural differences 
between the proteins which, however, alter the topology of the backbone, 
result in dramatic changes in their mechanical properties and stability. 
This research reveals that there is a strong relationships 
between the topological properties and functional features of biomolecules.

\section{Materials}

\subsection{Coarse-grained model}The coarse-grained molecular dynamics modeling
we use is described in detail in refs. \cite{Cieplak_2004,Cieplak_2004a,Sulkowska_2007b}.
In particular, the native contacts  between
the C$^{\alpha}$ atoms in amino acids $i$ and $j$,
separated by the distance $r_{ij}$,
are described by the Lennard-Jones potential
$V_{LJ} = 4\epsilon \left[ \left( \sigma_{ij}/r_{ij}
\right)^{12}-\left(\sigma_{ij}/r_{ij}\right)^6\right] \;$.  The length
parameter $\sigma _{ij}$ is determined pair-by-pair so that the minimum
in the potential corresponds to the native distance.
The energy parameter $\epsilon$ is taken to be uniform. As discussed in
ref. \cite{Sulkowska_2008d}, other choices for the energy scale and the form
of the potential are either comparable or worse when tested against
experimental data on stretching.
Folding is usually optimal at temperature $k_BT/\epsilon$ around 0.3
($k_B$ is the Boltzmann constant) which will be assumed as playing the role
of an approximate room temperature. Implicit solvent features come through
the velocity dependent damping and Langevin thermal fluctuation in the force.
We consider the overdamped situation which makes the characteristic
time scale, $\tau$, to be controlled by diffusion and not by ballistic-motion,
making it of order of a ns instead of a ps.
The analysis of the knot-related characteristics is made along the lines
described in ref. \cite{Sulkowska_2008}.

\subsection{KMT algorithm}
We determine the sequential extension of a knot, i.e.
the minimal segment of amino acids that can be identified as a knot,
by using KMT algorithm \cite{km1}. It involves removing the
C$^{\alpha}$ atoms, one at a time, as long as the backbone does not intersect a
triangle set by the atom under consideration and its two immediate sequential
neighbors.

\bigskip

\bigskip

\centerline{\Large{\bf Appendices}}

\appendix

\section{The two proteins}

In the classical arginine biosynthetic pathway, the OTCase enzyme
first deacetylates N-acetylornithine to the L-orinithine
and then forms citrulline through carbamylation.
In the other pathway involving AOTCase,
the process is reversed: N-acetylorintine is first
carbamylated to the N-acetylocitruline and then deactylated to the cirruline.

The folding, thermal, and mechanical properties of these two proteins
have not been compared up to now, mostly because the structure of AOTCase
has not been known until recently and because the presence of the knot makes
experiments harder to interpret.
However, it has been determined that both substrates N-acetylcitrulline (1yh1) and  L-
norvaline (1c9y) obey Michaelis-Menten kinetics \cite{Morizono}. It has also been
found \cite{Morizono} that affinity of the human OTCase to ornithine
is 10 times greater than the affinity of AOTCase to N-acetyl-ornithine,
while the affinity for carbamyl phosphate is approximately five times smaller.
The thermal stability was measured only for the OTCase 1c9y
(by measuring the temperature at which 50\% of the enzyme activity is lost)
was determined to be 56$\pm$1 $^o$C \cite{Morizono_1997}.

It has to be noted that there exists another member of the transcarbamoylase family,
SOTCase (extracted from B. Fragilis argF', 1sj1), which
also contains a knot. The rmsd between two structures, based on 280 equivalent
$C_{\alpha}$ positions, is around 1.4 \AA $\ $ \cite{Shi_1yh1}.
There are also structures which are similar
to the human OTCase 1c9y like E. coli ATCase (PDB code 1ekx) with
RMSD 1.7 \AA $\ $ based on 262 equivalent $C_{\alpha}$ atoms \cite{Shi_1yh1}.
The superpositions of these four proteins with their substrates
have been carried out in \cite{Shi_2006}, where only slight differences between
corresponding pairs of the knotted and unknotted proteins were found.
We have also checked that the properties of 1js1 and 1ekx in the model are nearly
identical as those of 1yh1 and 1c9y, respectively. For this reason, our analysis is
focused on the 1yh1 and 1c9y proteins.

%
%

\section{Typical trajectory of a knot in 1yh1 under stretching by a constant force}

We now analyze an example of the constant force pathway
 for 1yh1 for $F > 1.9$ $\epsilon$/{\AA} in more detail
({\it cf.} Figure 5 in the main paper and Fig. \ref{fig-cforce-knot-bis} below).
The right end of the knot is initially located close to
$\beta$-strand L (248-252).
The left end of the knot is located at the sharp turn involving glicyne
(170), and it is rather hard to force this end to leave this turn.
For this reason, when a constant force is applied,
the right end of the knot starts to move: as the protein backbone
is being pulled out of the loop, the sequential location of this
end decreases. Interestingly, the decrease is linear in time
(the center part in Figure 5) and it involves motion
of the right end across several pinning centers comprising
the turn at His-237, located between $\beta$-strand K (232-236),
the small helix H2
(238-244), and the sharp turn at Pro-210.
These centers are seen as accumulations of points along the tilted interval in
Figure 5. Nonetheless, these pinning sites are weaker than the force that
holds the left end at site 170 and hence it is only the right end that can slide.
The situation changes when the right end reaches Gly-200  within a sharp turn
at the end of a helix H1.
Interestingly, the left end is now ''pushed'' out of site 170 by the  right end
and it starts to move to the left, expanding the knot region a bit and resulting
in a translation of the whole knot to the left.
This translational motion stops on the left at Gly-164 at the end of the long
helix (147-164). At the same time the right end passes through
a half-loop (with a sharp and rigid turn involving prolines at 181 and 183)
between $\beta$-strand G (172-177) and helix H1 (185-199).
Finally, the right end stops at site 175 and the knot
becomes fully tightened. This scenario of events is consistent with the observation
of the role of the sharp structural turns in the dynamics of knot's
ends made in ref. \cite{Sulkowska_2008}.

We enclose a video presentations of the stretching of the two proteins,
as generated using our implementation of the Go-like model.
The first animation presents the protein 1c9y, and the second --
protein 1yh1. The process of tightening of the knot in the
animation corresponds to the Figure 5 of the main text. The knotted
region in 1yh1 and corresponding region in 1c9y are marked in green.

\section{Threshold force $F^*$}

We define $F^*$ as the threshold force at which the free energy barrier for the transition
from the native to the unfolded state vanishes and the protein begins to unfold in a
downhill manner. Unfolding is then essentially immediate,  without any intermediate states (
see the inset in Figure 7 in the main text).
The force $F^*$ is analogous to that found in simulations
of ubiquitin \cite{Szymczak_2006} above which the unfolding times are short and
distributed log-normally
and below which they are substantially longer and distributed exponentially.
For forces below $F^{*}$, the median
unfolding times follow a trend, which in general is a
 superposition of exponential functions \cite{Szymczak_2006}.
For forces above $F^{*}$, unfolding  times also decrease with an increasing
force, but at a much slower rate.

For 1yh1 and 1c9y we find $F^*$  of
$3.2$ and 2.5 $\epsilon/${\AA} respectively, as
indicated in figure 7 in the main text by the arrows.
The data shown in this figure are based on
300 trajectories for $F<1.9\epsilon/{\AA}$ and 100
trajectories for $F>1.9\epsilon/{\AA}$.
The relative shift in the location of $F^*$ is notable: $F^*$ for the knotted
1yh1 is higher pointing to a higher stability.

\section{Thermal untying of the knot in the 1yh1 protein.}

Fig. 9 in the main text shows the knot untying process in a schematic way.
Fig. \ref{knot_rozwijanie-bis} above shows the corresponding conformations
in more detail, with the original position of the knot along the backbone marked.

\section{Backtracking}

The process which involves a series of breaking and forming of the same group of contacts
due to topological barrier is called backtracking \cite{frust},\cite{Gosavi_2008}.
Complete thermal unfolding of the knotted proteins (i.e. unfolding to the trivial topology,
with the knot untightened) would not be possible without such backtracking.
An example of a backtracking due to knot topology is untying of the protein 1yh1 from the N
terminal. In this case the knot has to move along almost entire chain.
The translocation of the knot across the backbone is correlated with refolding due to
backtracking of a part of the structure, as seen in Fig. \ref{track}.
The bottom panel in that figure shows the number of contacts $Q$
in domain $b$ during unfolding. The top panel shows the number of contacts $Q$ inside the
domain $a$.
In the native state the position of the knot is stabilized by contacts G+I and I+K (in
domain $b$).
These contacts periodically break (black line), however until 2800$\tau$
the knot is localized in domain $b$, while in domain $a$ all contacts keep breaking
randomly.
When the knot moves to domain $a$ at  2800$\tau$,
the periodic refolding of contacts A+E is observed (top panel).
Eventually, the knot slides off the chain through the terminus N.


\bigskip

\bigskip

\centerline{\Large{\bf Acknowledgments}}

\bigskip

We thank D. Gront for help with reconstruction the proteins.
We appreciate discussions with D. Elbaum and P. Virnau.
This research has been supported by the grant N N202 0852 33 of the Ministry of
Science and Higher Education in Poland, by the grant PHY-0216576 and 0225630 from the
National Science Foundation (NSF)-sponsored Center for Theoretical Biological Physics.
P. Su{\l}kowski acknowledges the assistance of the Humboldt Fellowship,
as well as the hospitality of the University of California San Diego,
where a part of this project was done.

\newpage


\begin{table*}
\begin{center}
\tiny{
\begin{tabular}{c|c|c|c|c|c|c||c}
& \multicolumn{6}{|c||}{Constant velocity} & \multicolumn{1}{c} { Constant force} \\
& \multicolumn{6}{|c||}{} & \multicolumn{1}{c} { $F> 1.9 \epsilon$/{\AA}} \\
\hline
Order & 1yh1(typical)  &  1yh1(rare) & 1yh1$^{*}$(typ.) & 1yh1$^{*}$(rare) & 1c9y & 1c9y(with 
SS) & 1yh1 \& 1c9y  \\ 
\hline
1. & L+M (b)         & L+M (b)              & L+M        & L+M   &  L+M        & L+M  & L+M               
\\
2. & A+E (a)         & G+L (b)              & G+L        & A+E   &  G+L        & A+B  & 
G+L,I+K,G+I   \\
3. & C+D (a)         & G+I, H1,H2 (b)       & G+I,H1,H2  & C+D   &  I+K,H1,H2  & A+E  & A+E, E+F        
\\
4. & E+F (a)         & I+K (b)              & I+K        & E+F   &  G+I        & E+F  & A+B             
\\
5. & A+B (a)         & A+B (a)              & A+E,E+F    & A+B   &  E+F        & G+L, G+I,H1,H2  
&    \\
6. & G+L,I+K,H1,H2(b)& A+E,C+D (a)          & A+E,C+D    & G+L,I+K,H1,H2   &  A+E        & I+K             
&      \\
7. & G+I (b)         & E+F (a)              & A+B        & G+I             &  A+B        &                 
&   \\
\end{tabular}
}
\end{center}

\caption{The order of the contact breaking for different pathways}
\vspace{0.5cm}
\label{tab-F-d}
\end{table*}


\begin{figure}[htb]
\begin{center}
\includegraphics[width=0.52\textwidth]{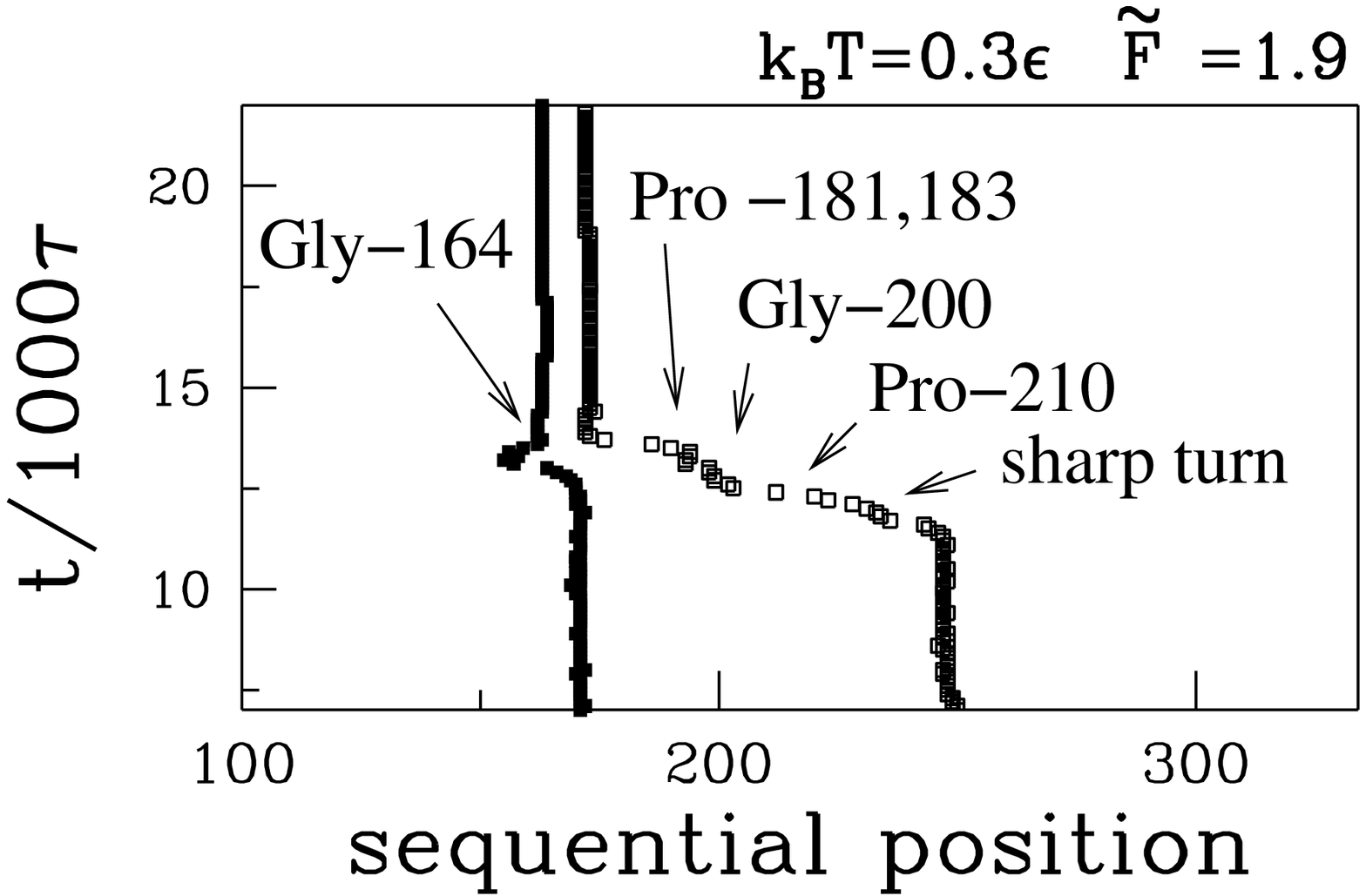}
\includegraphics[width=0.37\textwidth]{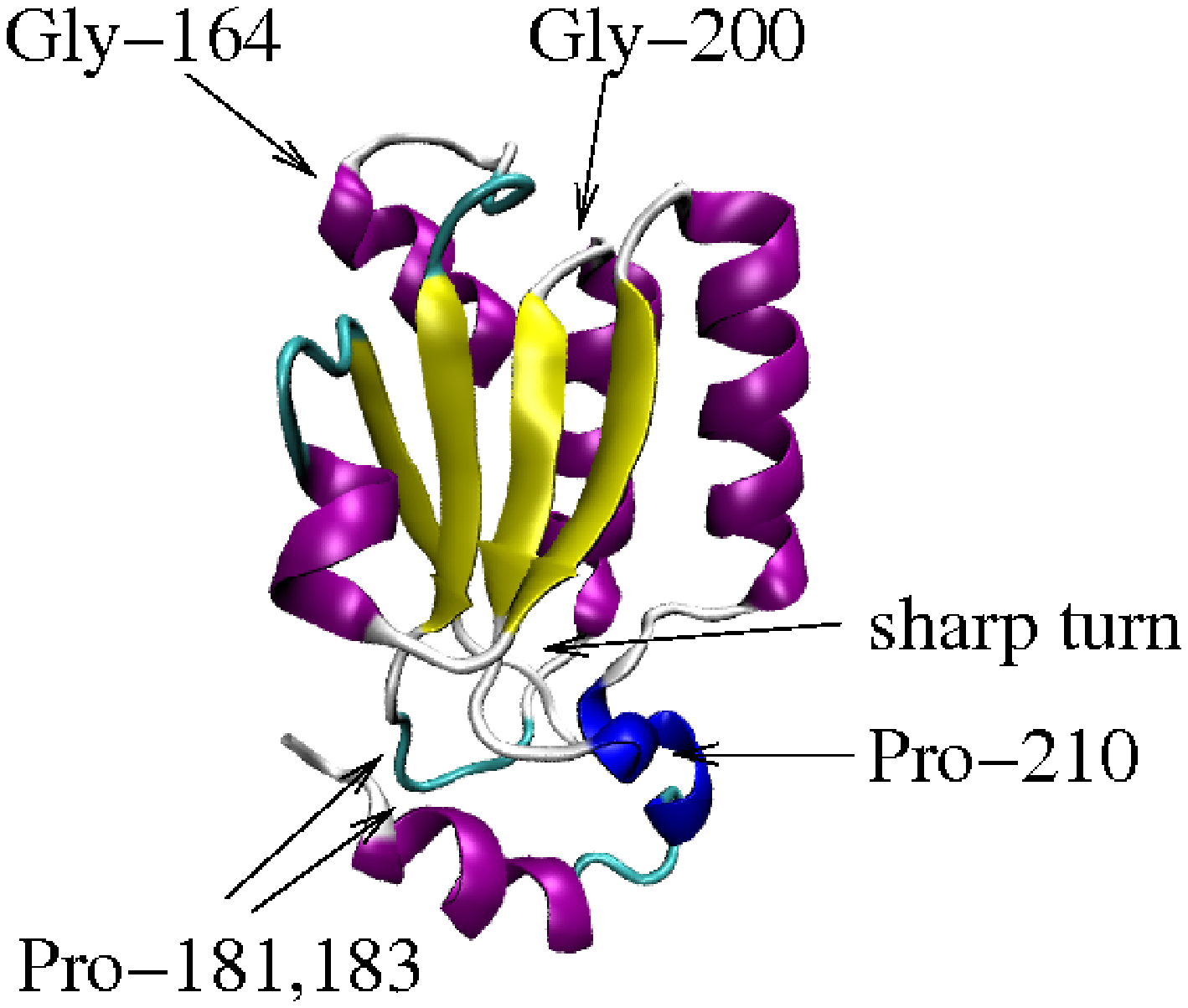}
\caption{Left, a typical trajectory of knot's ends locations for stretching at constant
force.
Right: the corresponding region of the knot in 1yh1 shown in the cartoon representation.
The pinning centers are indicated in the both panels.}
\label{fig-cforce-knot-bis}
\end{center}
\end{figure}

\begin{figure}[htb]
\begin{center}
\includegraphics[width=0.3\textwidth]{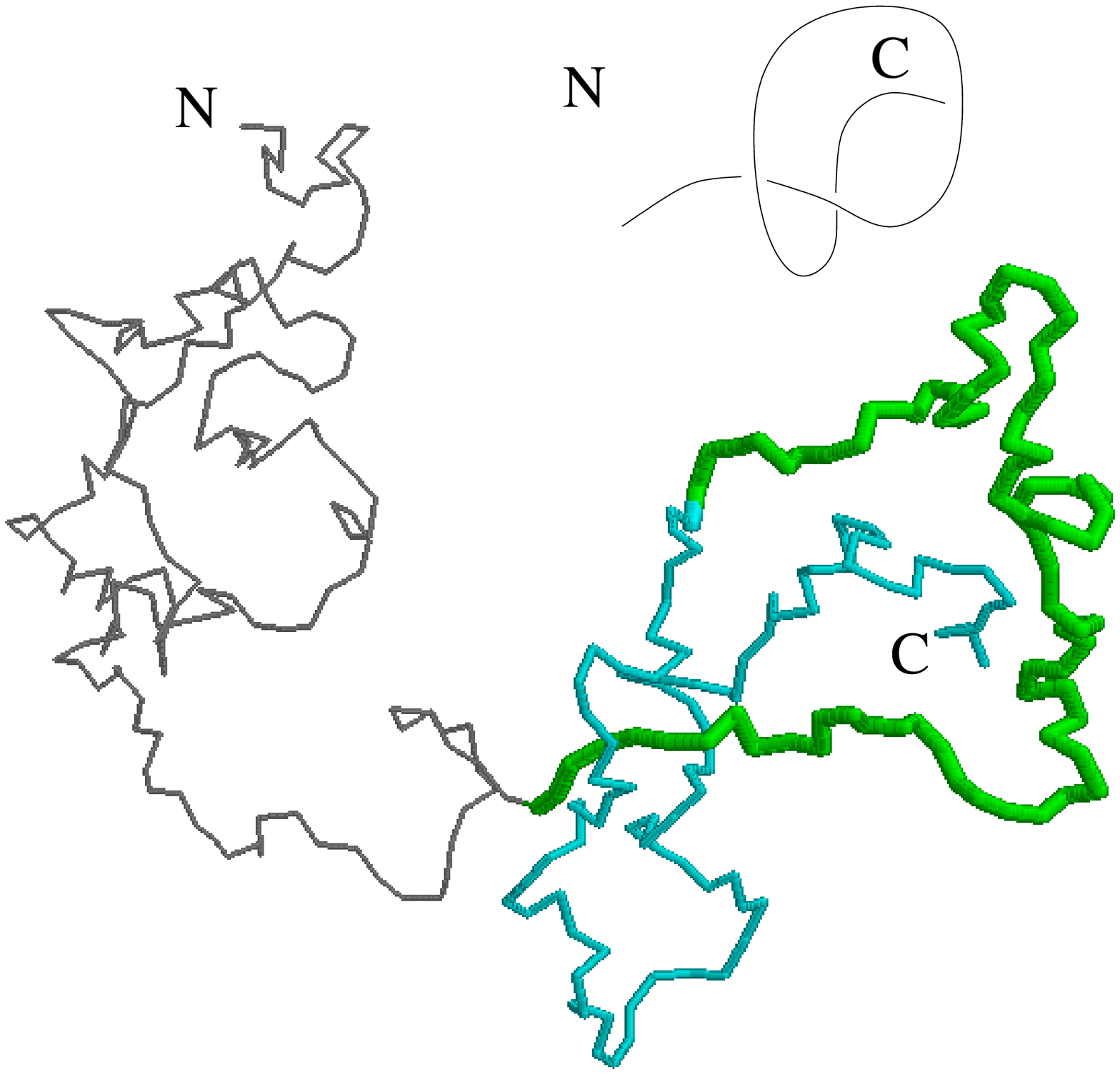}
\includegraphics[width=0.3\textwidth]{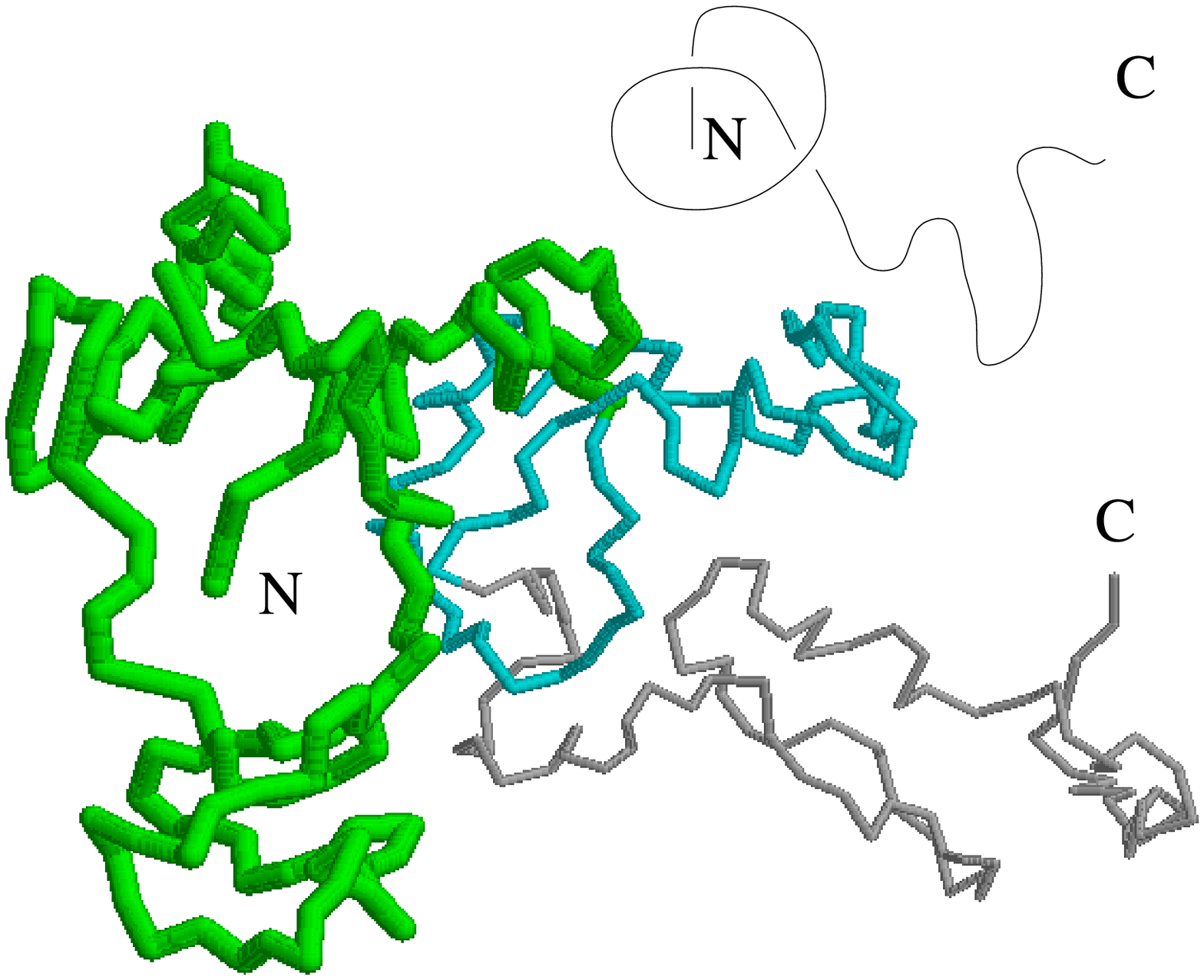}
\includegraphics[width=0.3\textwidth]{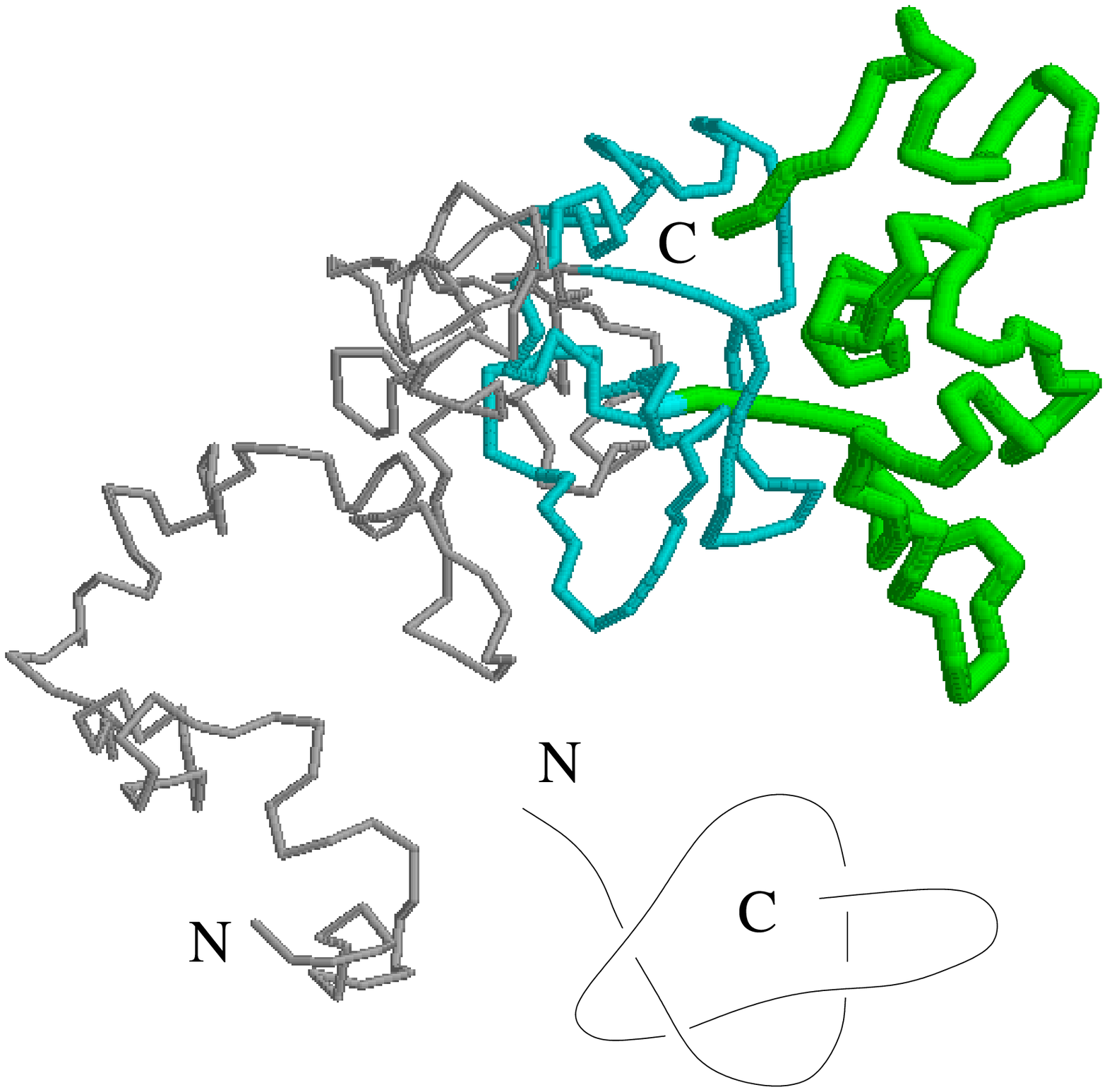}
\caption{Three possible ways of thermal untying of the knot. From the left to
the right: simple from the C terminus, simple from the N terminus
and through formation of a slipknot. The cyan (medium thick) line
indicates the native location of the knot, whereas the green (thick) line in the center and
left panel shows the instantaneous position of the knot. In the right panel,
the position of the slipknot is indicated by the combined
lines of medium and large thickness.} \label{knot_rozwijanie-bis}
\end{center}
\end{figure}

\begin{figure}[htb]
\begin{center}
\includegraphics[width=0.6\textwidth]{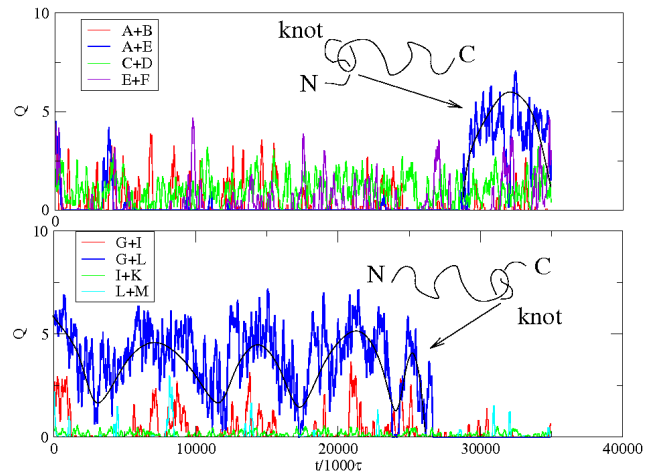}
\caption{Thermal untying of the protein accompanied by backtracking.
The bottom and top panels show respectively the number of contacts $Q$ in domains $b$ and
$a$ during unfolding.
The black line approximates periodic breaking of contacts G+I and I+K in the first phase of
unfolding, when the knot is still localized in domain $b$.
The knot moves from domain $b$ to $a$ around 2800$\tau$,
and eventually slides off the chain through the terminus N.
} \label{track}
\end{center}
\end{figure}

\end{document}